# End-to-end User Recognition using Touchscreen Biometrics

Michał Krzemiński [1*], Javier Hernando [2]

[1] Faculty of Electronics and Information Technology, Warsaw University of Technology, Poland
[2] TALP Research Center, Universitat Politecnica de Catalunya, Spain
[*] micha.krzeminski@gmail.com

**Abstract:** We study the touchscreen data as behavioural biometrics. The goal was to create an end-to-end system that can transparently identify users using raw data from mobile devices. The touchscreen biometrics was researched only few times in series of works with disparity in used methodology and databases. In the proposed system data from the touchscreen goes directly, without any processing, to the input of a deep neural network, which is able to decide on the identity of the user. No hand-crafted features are used. The implemented classification algorithm tries to find patterns by its own from raw data. The achieved results show that the proposed deep model is sufficient enough for the given identification task. The performed tests indicate high accuracy of user identification and better EER results compared to state of the art systems. The best result achieved by our system is 0.65% EER.

## 1. Introduction

Nowadays, mobile devices have become an essential element of modern society with almost three billion devices in use [1]. Smartphones are used not only for communication, but also for entertainment and work. Mobile devices are full of data about the user, starting from confidential emails, social media activity and ending with access to a bank account.

The most popular way of user's recognition used in smartphones is PIN or lock pattern. Both methods can be easily noticed and spied [2]. Therefore, biometric recognition, based on physical or behavioral characteristics, is increasingly used to secure mobile devices. Furthermore, using biometric data there is no need to remember passwords.

An example of physical biometrics is fingerprint. Fingerprint biometrics is well-researched and successfully implemented method of recognition in modern smartphones. It provides very high accuracy, but it is just one entry-point method, that is, the method cannot detect intruders after the identification step is performed successfully.

Behavioral biometrics provides the possibility of continuous, transparent user identification on mobile devices. Behavioral biometrics is based on measurements from actions performed by the user, not directly on physical features. An example of behavioral biometrics is touchscreen biometrics, which is the subject of this work.

User identification using touchscreen of mobile devices can be done using handwritten biometrics or touchscreen biometrics by using only simple traits and not characters. Handwritten biometrics require user to write a certain text, from which features are extracted. This method can be used in high security applications. Examples of handwritten biometrics system from touchscreen data are described in [3] [4]. However, in our work we consider touchscreen without handwritten biometrics, but based on natural user's touchscreen activity. The advantage of touchscreen biometrics is transparency. User is not aware of the fact of identification.

The touchscreen biometrics is based on human interaction with the touchscreen on mobile devices, by measuring shapes and strokes made by the user with the finger on it. Stroke is a sequence of consecutive time points, represented as screen coordinates. Biometric data is collected directly from the screen, and no additional sensors are required.

Interacting with the touchscreen is a key functionality of mobile devices. Touchscreen data is easy to collect and is allowed by default for all mobile applications [5]. Collecting data from the screen is transparent, non-intrusive to the user, does not affect his/her operation. Besides, collecting a large database of touchscreen biometric data, enables users' recognition cross devices and tracking their actions.

Furthermore, touchscreen biometrics permits Continuous Authentication (CA), a process in which the identity is constantly verified based on the activity of current user operating on the device [6]. When there is a doubt about the authenticity of the user, the system can block access to the device.

One of the first and most comprehensive articles in which touchscreen data is used as behavioral biometrics is [7]. In this study, 41 users provide data from touch operations while reading texts and comparing images. Only vertical and horizontal touch strokes are used. The system extracts 30 different features from the raw touchscreen data. The classification algorithms considered in the study are based on Support Vector Machine (SVM) with Radial-Basis Functions (RBF) and k-Nearest-Neighbors (kNN). The results presented in the study show that the combination of multiple strokes results in better performance. When deciding only with a single stroke, the EER is around 13%. Both classifiers obtain a lower error when increasing the number of strokes used in the classification. At the level of 11 to 12 strokes, the EER converges to the range between 2% and 3%.

In [8] 10 state-of-the-art touch-based authentication classification algorithms are compared under a common experimental protocol. The database consists of data from 190 users. As in the previous study, only vertical and horizontal touch strokes are used and 28 features are extracted. All 10 verification algorithms resulted in EER more than 10%. One of the tested algorithms is Multilayer Perceptron neural network, whose EER was around 14%.

In [9] system for active user authentication using touchscreen biometrics is presented. The authors perform user authentication using Kernel Sparse Representation-based classification (KSRC) and Kernel Dictionary-based Touch Gesture Recognition (KDTGR). These methods were compared for three datasets, two of them are public datasets mentioned in the paragraphs above, described in works [7] and [8]. The EER results range from 0.4% to 23.5%, depending on the dataset, the method and amount of strokes.

The aim in [10] is to assess the use of touchscreen biometrics in a continuous authentication task. In this study, a database with 71 users is used, and only 15 features are extracted. Two different neural networks are used for classification: artificial neural network (ANN) and Counter Propagation artificial neural network (CPANN). In this work, the neural network is not used directly for identification, but to update the value of the trust. When the trust value decreases below a threshold, the system decides that the current user is an impostor.

In [11] an analysis of touch-interaction behavior for active user authentication was performed using SVM, kNN, Random forest and Neural Networks. EER results achieved are between 1.72% and 9.01%. The authors concluded that the authentication accuracy improves with higher number of strokes used in the authentication or small timespan between the training and testing phases.

In [12] three methods for user authentication based on SVM and Gaussian Mixture Models (GMM) with handcrafted features were compared on four public touchscreen datasets. Two databases were mentioned before, described in works [7] and [8]. Other database consisted of data from 71 users captured from 8 devices with different screen size, described in [13]. The last database consisted of data from 48 users [14]. The authors in [12] processed independently strokes of different orientation (down, up, left, right). The achieved EER results range from 3.1% to 27.8% for 10 strokes.

All related works on touchscreen biometrics are based on hand-crafted features, by using different classification algorithms. In this paper we consider touchscreen biometrics, which provides transparency to user identification. It uses data from natural user's touchscreen activity, simple horizontal and vertical gestures.

The original contribution of our work is proposing first end-to-end user identification system using touchscreen biometrics. The implemented deep learning classification algorithm does not use hand-crafted features, but tries to find patterns by its own from raw touchscreen data, extracted directly from touchscreen sensor.

The rest of this paper is summarized as follows. Section 2 describes our proposed user identification method based on touchscreen biometrics. Section 3 describes implementation of our end-to-end system, touchscreen database and experimental setup. Sections 4 shows the evaluation of performance of created recognition system. We conclude this paper in Section 5.

## 2. Proposed user identification method

The task of the proposed biometric system is user's recognition based on data from the touchscreen. User interactions with the touchscreen are recorded using the screen itself, so no additional sensors are needed [15], and only vertical and horizontal gestures were selected due to the distinctive characteristic of these gestures [7]. The vertical gesture corresponds to scrolling the screen up and down, the horizontal gesture corresponds to flipping the screen from side to side.

Most biometric systems perform feature extraction from used biometric data. These features are hand-crafted features designed beforehand by human experts. Then these features are used as input to the classification algorithm. However, in this work we use an end-to-end deep neural network whose input are the raw data from touchscreen sensors. No hand-crafted features have been used. Deep learning models must learn patterns from raw data, which may result in finding new, previously unknown data characteristics.

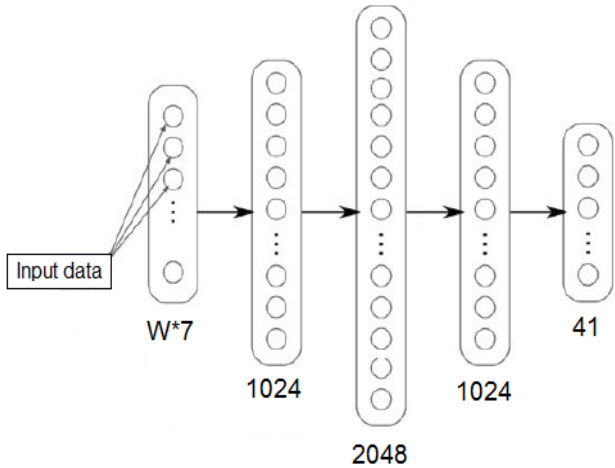

*Fig. 1. Deep neural network architecture. Input layer size depends on the window size (W)*

The selected deep learning model in the proposed system is a deep multilayer perceptron (MLP). The MLP has been chosen because of its simplicity. The architecture of the network is presented in Fig. 1. The proposed deep network is composed of three hidden layers with a high number of neurons in order to provide sufficient capacity to represent the complex function of touchscreen strokes, and the output layer has as many neurons as users in the database. Regarding the input layer, the problem is that the strokes can have different lengths, but MLP networks are characterized by a fixed number of neurons in the input layer. To overcome this issue, a sliding window has been used. Only the windowed part of the stroke, is shown to the network input and the classification is based on it. The number of input layer neurons is the number of biometric attributes from the database multiplied by the window size, *W* in Figure 1. The sliding window mechanism uses raw biometric data from the database and preserves the temporality of strokes.

## 3. Experiments

### 3.1. Database

The proposed system uses a public touchscreen database [16]. The procedure for collecting data for this database was described in the protocol [7], in which respondents were asked to read three text documents on Wikipedia on a mobile device and, after each document, to answer questions about understanding the text. In this way, data about vertical gestures was obtained. The second phase of the protocol consisted of asking users to detect differences on pairs of similar images. This process involved the need to perform horizontal gestures in order to flip the screen between the pictures. In addition, one more document and one more pair of images were collected one week after the first study in order to collect biometric data with a time gap between them. During the data collection process, four different mobile devices were used:
- Nexus 1,
- Nexus S,
- Samsung Galaxy S,
- Droid Incredible.

All user touchscreen data was collected in *csv* format with more than 900k records. Each record contains a touch sample with 11 attributes:
- phoneID - ID of used mobile device,
- userID - ID of tested user,
- documentID - ID of document used in test,
- timestamp,
- action - starting/stopping touching screen,
- phone orientation,
- x-coordinate,
- y-coordinate,
- pressure,
- area covered - by finger on screen,
- finger orientation.

Not all attributes present biometric data. That is why in our system we do not use the *phoneID*, *documentID* and *timestamp* fields to train the neural network. The *userID* field is used as a label to train the network in supervised manner.

Each database record is a time sample. Initial preprocessing of the database was done to group records into strokes. Each stroke starts when the user touches the screen (action field equal to 0) and ends when the user lifts the finger (action field equal to 1). Each stroke can have different lengths.

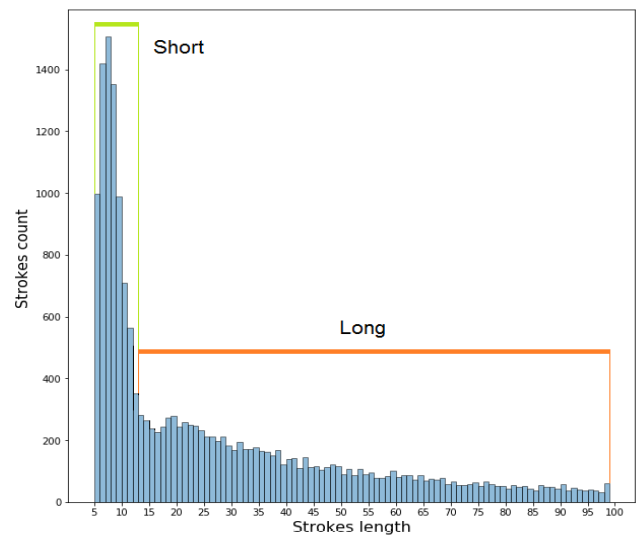

*Fig. 2. Stroke length histogram*

The histogram of gestures in terms of their length is shown in Fig. 2. To filter the database from unwanted strokes, we flagged too short strokes (having less than 5 touch samples) as outliers and discarded them before proceeding to the learning phase, as they are not useful for our task. When analyzing the histogram, the strokes were classified in two groups: short and long. Short strokes are gestures with a length in the range of 5 to 12 records, long strokes are sequences over 13 records. In this study, the system was tested using two different database sets:
- using all strokes for learning,
- using only long strokes.

The number of strokes taken into learning has a strong impact on learning time.

Our network deal with classification task, that is why number of output neurons is equal to users in database. The database is imbalanced. The number of strokes is different per each user. This fact has a strong impact in the learning of the neural network. That is why we calculated class weights as amount of each user strokes in relation to all strokes and then used class weights in the learning process.

The database used in this study was divided into three parts. 60% of the database is used for deep network learning, 20% for network validation and 20% is used as a test set for final network evaluation. Strokes were divided into each of these parts to have a proportional number of strokes for each user in each part. There are over 20k strokes in the database.

A main difficulty in our network is overfitting [17]. The overfitting prevention mechanism used is 50% dropout applied after the first hidden layer and batch normalization. The effect of this was a reduction of validation loss. The validation error stabilized with subsequent epochs.

*3.2. Implementation and assessment*

The classification algorithm has been implemented using Keras framework [18] and Google Colaboratory environment [19]. Learning and testing deep neural networks with a large number of parameters requires large computing resources and a large amount of time [20]. By implementing the system in the colab notebook, it transfers the processing into the cloud, where hardware acceleration on GPU could be used.

The implemented neural network takes a decision for user recognition based on only part of a stroke restricted by a window. In order to take a decision based on the entire stroke, the scores for multiple windows were fused, which resulted in significant increase in system accuracy [21]. A further fusion of multiple strokes into decision was performed and tested. For each number of strokes used in fusion, FAR (false acceptance rate) and FRR (false rejection rate) were calculated as the number of incorrectly accepted impostors and the number of incorrectly rejected users, using equations (1) and (2) [22]. Where *FP* is number of false positives, *FN* is number of false negatives, *TN* is number of true negatives and *TP* is number of true positives.

$$FAR = \frac{FP}{FP + TN} \quad (1)$$

$$FRR = \frac{FN}{FN + TP} \quad (2)$$

Finally, the EER (equal error rate) was obtained fixing the verification threshold, which yields the same FAR and FRR. The system results and performance will be described in next section.

## 4. Results

The initial result of deep network learning when making decision based on only part of the stroke, restricted by a window of size 5, resulted in only 65% classification success. As mentioned before, in order to improve the accuracy, the fusion of all windows of individual strokes were made. A further fusion of windows for multiple strokes of the same user also were tested. These results are presented in Fig. 3, where the horizontal variable is the number of fused gestures and the vertical variable is the accuracy. The diagram shows two curves corresponding to the type of strokes from the database used. In case of all strokes, after fusing one stroke, the accuracy increases from 65% to 77.89%. Further fusion increases accuracy up to 91.62% for 10 strokes.

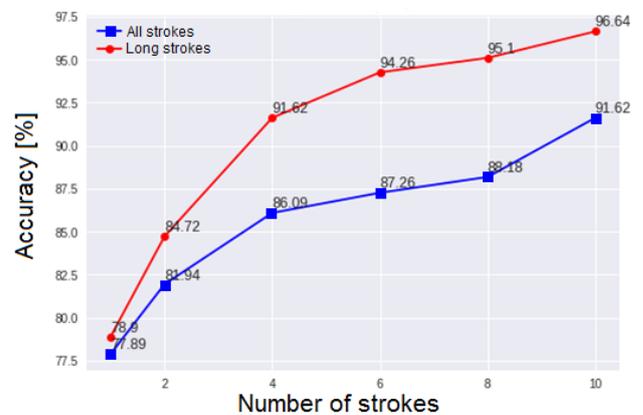

*Fig. 3. System accuracy in terms of number of strokes*

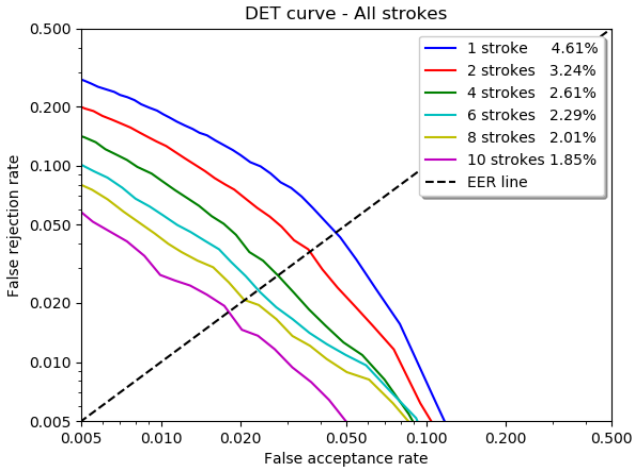

*Fig. 4. Detection error tradeoff curves for all strokes*

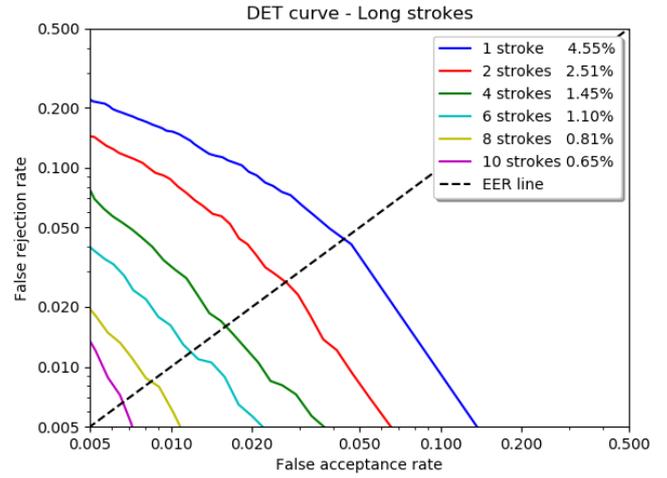

*Fig. 5. Detection error tradeoff curves for only long strokes*

In the case of user recognition based on only long strokes, the accuracy for long strokes is better than for all strokes. This is expected behavior, because long strokes contain more biometric information. The best accuracy achieved by the proposed system is 96.64%, making fusion of 10 long strokes.

In order to compare the proposed system with previous works, the EER metrics has been calculated for the both cases considered in this work. Detection error tradeoff (DET) curves in terms of the amounts of the strokes are presented in Fig. 4 for all strokes, and in Fig. 5 for only long strokes. EER values for each amount of strokes are presented on the legend of the figures. The EER when using only one stroke is 4.61%, the EER value decreases by adding more strokes, until 1.85% for 10 strokes. The results in the case using only long strokes are shown in Fig. 5, that show lower error than in the case when all strokes are used for the recognition. The best EER result achieved by the created system is 0.65% EER fusing 10 long strokes.

A comparison of the proposed system with touchscreen biometric systems [7], [12] and [9] has been carried out and is shown in Table 1. Those results are comparable, because they used the same database, described in [7]. The EER achieved in Frank's publication [7] for one stroke is 13%. The error decreases with greater amount of strokes and stabilizes at 3% EER for 10 strokes. In [9] two methods were used. KSRC (Kernel Sparse Representation-based classification) achieves 17.6% EER for one stroke. The error decreases to 1.31% EER for 9 strokes fused. Better results has been achieved by KDTGR (Kernel Dictionary-based Touch Gesture Recognition) method with 8.51% EER for one stroke and 0.6% for 9 strokes.

Our proposed system has achieved 4.61% EER for one stroke decreasing to 1.85% EER for 10 strokes. Those results show that our proposed method outperform the SVM classifier. Comparing to results in [9], our system achieves better results on smaller amount of strokes (up to 5 strokes). On larger amount of strokes, results in [9] are better.

| Strokes | Proposed method | SVM [7] | KSRC [9] | KDTGR [9] |
|---|---|---|---|---|
| 1 | 4.61 | 13 | 17.6 | 8.51 |
| 2 | 3.24 | 8.5 | - | - |
| 3 | 2.91 | - | 6.13 | 4.05 |
| 4 | 2.61 | 6.5 | - | - |
| 5 | 2.32 | - | 3.42 | 2.29 |
| 6 | 2.29 | 5 | - | - |
| 7 | 2.12 | - | 2.19 | 1.14 |
| 8 | 2.01 | 4 | - | - |
| 9 | 1.92 | - | 1.31 | 0.6 |
| 10 | 1.5 | 3 | - | - |

*Table 1 EER comparison between proposed system and state-of-the-art systems [7] [9]*

The comparison between our proposed system and [12] is presented in Table 2. The best EER result achieved in [12] was fusing SVM and GMM methods, 3.1% for 10 strokes. On the other hand our method based on Deep Neural Network achieved better results with 1.85% EER for all strokes and 0.65% EER for only long strokes.

The main difference in the approaches between our proposed system and other state-of-the-art studies is the use of a different classification algorithm for identification. All those methods used hand-crafted features extracted from the touchscreen data. On the contrary our approach uses a deep neural network whose inputs are the raw data from the touchscreen. The presented results show that a deep neural network can be successfully used in identification task and can outperform state-of-the-art methods.

| Method | SVM | GMM | Fused | Proposed method |
|---|---|---|---|---|
| EER [%] | 5.8 | 9.1 | 3.1 | 1.85 all strokes<br>0.65 long strokes |

*Table 2 EER comparison between proposed system and system described in [12]*

Our results show that the implemented deep neural network with three hidden layers is able to model a complex representation of the touchscreen data. The implemented network does not use hand-crafted features, but tries to find patterns by its own from raw data. This approach has not been considered before for touchscreen biometrics. This approach does not need experts to investigate and design discriminative features. The presented results show that a fully automatic end-to-end system with a deep neural network can surpass the accuracy of other systems.

The learning time for a deep neural network is long and requires a large database of touchscreen data, but this load is required only once in the learning phase of the system. After the learning phase, the recognition is fast enough for real time user identification.

## 5. Conclusion

This work has studied touchscreen data as behavioral biometrics. The aim of the study was to create an end-to-end system that can identify users from mobile devices using deep learning model. The implemented system uses a deep neural network that takes as input raw data and is able to decide on the identity of the user with high accuracy.

One main difference of the proposed method compared to other state-of-the-art systems is the feature extraction. The implemented classification algorithm does not use hand-crafted features, but tries to find patterns by its own from raw touchscreen data.

The deep neural network has been tested in two cases: using all strokes from the database and using only long strokes. Tests conducted in each of these cases show that the proposed deep architecture is sufficient for the given task. The fusion of strokes resulted in significant increase in system accuracy, being the best result achieved by the proposed system 96.64% and 0.65% EER for 10 long strokes. The accuracy achieved by the system that was learned using long strokes is better, because long strokes contain more biometric information.

Experimental results in this work show better EER results in comparison to state-of-the-art systems based on SVM and GMM methods. On the other hand our Deep Learning approach achieved better results than methods based on dictionaries for small amount of strokes. On large amount of strokes dictionaries methods achieved better results. Future works should consider a fusion between both approaches to achieve good performance for any amount of strokes.

The presented study proves that the user can be identified with high accuracy using data from the touchscreen of the mobile device. The touchscreen data is a key functionality of mobile devices and is allowed by default to every mobile application. Touchscreen biometrics is a transparent method of user identification, and can be used to improve mobile security by continuous authentication. In conclusion, this work shows that touchscreen biometrics is an important area for further research.